\documentclass[referee,psfig]{aa}
\usepackage{graphicx,natbib,longtable}
\bibpunct{(}{)}{;}{a}{}{,}
\begin{document}


\title{Extinctions at 7${\rm \mu}$m and 15${\rm \mu}$m from the ISOGAL survey}

\author{B.W. Jiang\inst{1},  J. Gao\inst{1}, A. Omont\inst{2},
F. Schuller\inst{2}, G. Simon\inst{3} }
\offprints{Biwei Jiang (bjiang@bnu.edu.cn)}

\institute{Department of Astronomy, Beijing Normal University,
Beijing 100875, P.R.China
\and
 Institut d'Astrophysique de Paris, CNRS, 98 bis Bd Arago, F-75014
 Paris, France
\and
GEPI, Observatoire de Paris, France}

\date{Received date: \today ,  Accepted date:}

\titlerunning{Extinction at 7${\rm \mu}$m and 15${\rm \mu}$m }
\authorrunning{Jiang et al.}

\abstract{ The extinction laws at 7${\rm \mu}$m and 15${\rm \mu}$m
are derived for more than 120 sightlines in the inner Galactic
plane based on the ISOGAL survey data and the near-infrared data
from DENIS and 2MASS. The tracers are the ISOGAL point sources
with [7]-[15]$<$0.4 which are RGB tip stars or early AGB stars
with moderate mass loss. They have well-defined intrinsic color
indices (J-Ks)$_0$, (Ks-[7])$_0$ and (Ks-[15])$_0$. By a linear
fitting of the observed color indices Ks-[7] and Ks-[15] to the
observed J-Ks, we obtain the ratio between the E(Ks-[7]) and
E(Ks-[15]) color excesses and E(J-Ks). We infer the selective
extinctions at 7 and 15${\rm \mu}$m in terms of the near-infrared
extinction in the Ks band.  The distribution of the derived
extinctions around 7 micron (A$_{7}$) is well represented by a
Gaussian function, with the peak at about 0.47A$_{\rm K_{S}}$ and
ranging from 0.33 to 0.55A$_{\rm K_{S}}$ (using the near-infrared
extinctions of Rieke \& Lebovsky 1985). There is some evidence
that A$_{\rm 7}$/A$_{\rm Ks}$ may vary significantly depending on
the line of sight. The derived selective extinction at 15${\rm
\mu}$m suffers uncertainty mainly from the dispersion in the
intrinsic color index (Ks-[15])$_0$ which is affected by dust
emission from mass-losing AGB stars. The peak value of A$_{\rm
15}$ is around 0.40A$_{\rm K_{S}}$.
 }

\maketitle

\section{Introduction}

In the mid-infrared wavelength range, the interstellar extinction
is basically composed of two components [see e.g.
\citet{Draine89}]. One is the continuum extinction decreasing with
wavelength which can be attributed to graphite or PAH carbonaceous
dust grains; the other relates to the spectral features around
10${\rm \mu}$m and 18${\rm \mu}$m due to silicate dust grains. The
extinction around 7${\rm \mu}$m is rather controversial, partly
due to the fact that this wavelength is where the decreasing
continuum extinction meets the rising silicate extinction peak.
Before ISO (Infrared Space Observatory) was launched, there were
several measurements of near-infrared and mid-infrared extinctions
which could be used to derive that at 7${\rm \mu}$m.
\citet{Rieke85} observed \emph{o} Sco and a number of stars in the
Galactic center and derived the extinction law in the 1-13${\rm
\mu}$m range; \citet{Whittet88} reobserved \emph{o} Sco;
\citet{Landini84} observed hydrogen recombination lines from the
obscured HII region G333.6-0.2. \citet{Draine89} concluded that
the available observational evidence was consistent with the power
law A$_{\lambda}\propto\lambda^{-1.75}$ out to
$\lambda\approx7\mu$m (which means the extinction at 7${\rm \mu}$m
is about 0.13A$_{\rm K}$) and that the various extinction curves
from different observations appear to be essentially universal for
wavelengths $\lambda>0.7\mu$m.

After the launch of ISO, new determinations were made of the
extinction around 7${\rm \mu}$m. Using ISO observations of
hydrogen recombination lines, \citet{Lutz96} and \citet{Lutz99}
found that the extinction toward Sgr A$^{*}$ does not decline with
increasing $\lambda$ in the 4-8${\rm \mu}$m region, and obtained
an extinction of about 0.45A$_{\rm K}$ at 6.8${\rm \mu}$m.
However, from their studies using H$_{2}$ rovibrational lines in
the Orion molecular cloud outflow, \citet{Rosenthal00} found that
the extinction continues to decline with increasing $\lambda$ to a
minimum at about 6.5${\rm \mu}$m, and the extinction at 7$\mu$m is
about 0.15A$_{\rm K}$, though their lower value may be attributed
to the relatively modest extinction of the field they analyzed,
A$_{\rm Ks}$ being only 1 mag. Moreover, by comparing the
intensity of infrared dark clouds and their neighboring emission
and assuming a simple radiative transfer model,
\citet{Hennebelle01} derived an average opacity radio of 7${\rm
\mu}$m to 15${\rm \mu}$m of 0.7$\pm$0.1. The later two results
from \citet{Rosenthal00} and \citet{Hennebelle01} are consistent
with the result by \cite{Draine89}. The issues of whether the
discrepancy between the results from \citet{Lutz99} and from
\citet{Rosenthal00} or \citet{Hennebelle01} is real or not and
whether it could result from dust properties that depend on the
line of sight need more investigation.

With data available from numerous lines of sight in the Galactic disk, the
ISOGAL and GLIMPSE \citep{Benjamin03,Churchwell04} projects provide an
opportunity to study the dependence of the extinction on the interstellar
dust environments.

The ISOGAL project is a survey of 263 fields, covering about 16 square
degree area in total, in the Galactic plane. It made use of the ISOCAM
camera aboard the ISO space telescope at 7${\rm \mu}$m and 15${\rm
\mu}$m  \citep{Omont03}. Furthermore, most of the ISOGAL
fields in the southern hemisphere were also observed by the DENIS project
\citep{Epchtein99}, in the IJK$_{\rm S}$ bands. The ISOGAL data were
systematically merged with DENIS in a five-wavelength catalog
\citep{Schuller03}. In the meantime, another ground-based near-infrared sky
survey project, 2MASS, published data covering almost the whole sky,
including the ISOGAL sky area. In combination with the DENIS and the 2MASS
data, ISOGAL has great potential for studying the extinctions at 7 and
15${\rm \mu}$m in the Galactic plane.

By combining J-Ks and Ks-[7] colors obtained from DENIS and ISOGAL
\citep{Omont03,Schuller03}, we \citep{Jiang03} derived the selective
extinction at 7 and 15${\rm \mu}$m along the line of sight towards an ISOGAL
field at \emph{l}=18.3 and \emph{b}=-0.35, obtaining A$_{7}$ $\sim$
0.4A$_{\rm Ks}$. Although this result suffers from some uncertainty, it is
more consistent with \citet{Lutz96} than with \citet{Rosenthal00}.

The most recent result concerning the mid-infrared extinction is
from Spitzer. The GLIMPSE program observed a large part of the
Galactic disk in four bands between 3.6 and 8$\mu$m
\citep{Indebetouw04}. \citet{Indebetouw04} found, along two lines
of sight, that the extinction across the 3--8$\mu$m wavelength
range is flat, which agrees with \citet{Lutz99}, as well as
ourselves \citep{Jiang03}.

Given that the wide areas covered by ISOGAL and, more
systematically, by GLIMPSE, include various sightlines (and thus
various extinction environments) one can check whether the mid-IR
extinction varies. The present paper aims at checking the
extinction at 7 and 15${\rm \mu}$m. We have used in principle the
same method as in \citet{Jiang03} to determine the extinction law
at 7 and 15${\rm \mu}$m from the ISOGAL fields. Since the ISOGAL
fields are within the Galactic plane and lie mostly in the inner
disk, this work mainly relates to the central parts of the
Galactic plane.

\section{Database}
\subsection{The ISOGAL PSC data}

This work is based on the ISOGAL-DENIS Point Source Catalog
Version 1.0 \citep{Schuller03} which includes more than 100,000
point sources detected at 7${\rm \mu}$m and/or 15${\rm \mu}$m in
the ISOGAL survey. In addition to the standard data reduction by
the CIA package \citep{Ott97} of the ISOCAM images, the point
sources are extracted by a dedicated PSF fitting procedure.
Because most of the ISOGAL survey areas are near the Galactic
plane ($|b|<1\degr$) and have high source density, the source
extraction is confusion limited and disturbed by strong background
radiation. The photometric error is usually better than 0.2mag and
rises above this value at the ISOGAL sensitivity limit of about
10mJy. The ISOGAL PSC catalog consists of three parts. The first
part includes 163 fields observed at both 7${\rm \mu}$m and
15${\rm \mu}$m, with fields denoted by ``FC". The second part
includes 43 fields observed only at 7${\rm \mu}$m, denoted by
``FA". The third part includes 57 fields observed only at 15${\rm
\mu}$m, denoted by ``FB".

Unfortunately, the ISOGAL observations did not use the same
filters through all the fields: towards about 75\% of them the
broad-band filters LW2 and LW3 were used, while towards the other
25\%, the LW5, LW6 and LW9 narrow-band filters were used
\citep{Schuller03}. The reference wavelengths of the LW2 and LW3
filters are 6.7${\rm \mu}$m with a bandwidth of 3.5${\rm \mu}$m
and 14.3${\rm \mu}$m with a bandwidth of 6.0${\rm \mu}$m,
respectively. The LW5, LW6 and LW9 filters have reference
wavelengths at 6.8${\rm \mu}$m, 7.7${\rm \mu}$m and 14.9${\rm
\mu}$m, with bandwidths of 0.5${\rm \mu}$m, 1.5${\rm \mu}$m and
2.0${\rm \mu}$m, respectively \citep{Blommaert03}. Thus, what we
call ``7${\rm \mu}$m'' or ``15${\rm \mu}$m'' is an approximation
and should not be confused with the precise central wavelengths of
the bands. Detailed qualitative information on the ISOGAL filters,
including the reference wavelengths, bandwidths, zero point
magnitudes and flux densities, and total observed area, can be
found in Table 1 of \citet{Schuller03}.

\subsection{The ISOGAL-DENIS PSC data}

The nominal position error is about 0.5\arcsec ~ in the DENIS survey, while
it is several arc seconds in the ISOGAL survey. Cross-identification with
DENIS has thus improved significantly the positional accuracy of the ISOGAL
point sources. The final position accuracy of the ISOGAL-DENIS PSC is
normally about 1 arc second. To deal with the confusion problem, the Ks
limiting magnitude was cut to give a source density of 72000/deg$^2$ for the
ISOGAL 3\arcsec pfov [pixel field-of-view] observations or 18000/deg$^2$ for
the ISOGAL 6\arcsec pfov observations.  The details of the ISOGAL-DENIS
catalog are described in \citet{Schuller03}. Although there are over 100,000
sources in the ISOGAL-DENIS PSC, only data of high quality are used in the
present work. We chose only sources whose association flags are 4 or 5,
implying that the cross identifications are reliable.  When there are two
sources within the search radius, the closest DENIS source is associated
with the ISOGAL source, and the association flag is usually reduced by 1
(from 5 as the highest quality). Such cases are mostly dropped from the
following analysis.  In addition, a constraint was also applied to the
photometric quality of the DENIS PSC sources. The correlation factors in
both J and K bands were required to be larger than 0.85. Since DENIS
surveyed only the southern sky, it did not cover all the ISOGAL fields. The
final number of retained associated sources is 50724 distributed in 220
fields, among which are 8552 sources in 42 FA fields (7$\mu$m only), 7107
sources in 56 FB fields (15$\mu$m only) and 35065 sources in 122 FC fields
(both 7 and 15$\mu$m).

\subsection{The ISOGAL-2MASS PSC data}

The Two-Micron All Sky Survey, (2MASS, \citet{Skrutskie97}) is
similar to DENIS in that it is also a ground-based near-infrared
survey. However, 2MASS differs from DENIS, mainly in that it
covers the whole sky including all the ISOGAL fields. In addition,
2MASS observed in the JHKs bands, i.e. including the H band but
not the I as in DENIS. The depth of the 2MASS photometry is
slightly greater than that of the DENIS photometry; nominally the
limiting magnitude for point sources is Ks$=$14.3mag compared to
14 mag in DENIS. As the field of view per pixel is similar, the
positional accuracy of the 2MASS PSC is of the same order as that
of DENIS.

The 2MASS PSC database was released in 2003 \citep{Cutri03}, after
the ISOGAL-DENIS PSC. The cross-identification procedure was the
same as for the ISOGAL-DENIS case, described in
\citet{Schuller03}. There are three major steps. First, we
determined the offsets of every ISOGAL field according to the
positions provided by the 2MASS PSC. As with all the southern
ISOGAL fields, this correction was accomplished by
cross-identification with the DENIS PSC. As the 2MASS astrometric
accuracy is comparable to that of DENIS (both have a nominal error
of 0.5\arcsec), this would not have been any significant
improvement to the positional accuracy. Consequently, this step is
applied only to the 42 northern fields which were beyond the DENIS
coverage. The offsets of these ISOGAL fields to the 2MASS PSC
sources range as high as 4 to 5 arcseconds though usually they are
less than 3\arcsec, similar to the values of the ISOGAL field
offsets to the DENIS PSC sources, as expected. After removing the
offsets, the positional accuracy of sources in the northern ISOGAL
fields is about 1.5\arcsec.  The second step is to search for the
2MASS counterpart of each ISOGAL point source. The search radius
is 1.8{\arcsec} or 3.5{\arcsec} for sources in the ISOGAL fields
with pfov of 3{\arcsec} and 6{\arcsec} respectively. If there is
more than one 2MASS source within the search radius, no 2MASS
source is assigned to the ISOGAL source (although the closest one
was taken to be the ISOGAL-DENIS cross-identification).  Also
rejected are the sources whose positional differences between
ISOGAL and 2MASS are bigger than 3$\sigma$, where $\sigma$ means
the rms ISOGAL-2MASS position difference in the corresponding
field (after field offset correction). The third step is to try to
avoid the confusion problem which was met with in the
cross-identification with the DENIS sources. It is likely to be
more serious because of the slightly higher sensitivity of 2MASS.
The 2MASS source density is limited to 72000/deg$^{2}$ or
18000/deg$^{2}$ for the ISOGAL fields with pfov 3\arcsec or
6\arcsec respectively, which corresponds to a formal random
association probability of 5\%, implying that the number of
spurious associations should be smaller than 1\%
\citep{Schuller03}. This step is accomplished by setting the
limiting magnitudes in the Ks band above the 2MASS limit (which
varies from field to field and ranges from about 12 to 14 mag).
The constraints on the association radii and confusion limits are
similar to good association quality flags (4 or 5) in the
ISOGAL-DENIS PSC. In addition to these steps, which are similar to
those involved in the ISOGAL-DENIS association process, the 2MASS
PSC sources are required to have reliable photometric results by
only accepting photometry quality flag ``AAA'' in the JHKs bands,
which means ${\rm S/N} >$ 10 in all bands. The read flags in all
three bands were also required to be good (i.e. ``Rflag'' being 1,
2 or 3), which indicates best quality detection, photometry and
astrometry \citep{Cutri03}. Although these two criteria may be
redundant in some sense, they are consistent and guarantee the
quality of data. Under these constraints, the resultant
ISOGAL-2MASS PSC contains 50653 sources distributed in all the 263
fields. These sources are composed of 8112 in 43 FA fields, 6623
in 57 FB fields and 35918 sources in 163 FC fields; or 39017
7${\rm \mu}$m sources in 205 fields and 23359 15${\rm \mu}$m
sources in 218 fields. Compared to the ISOGAL-DENIS PSC sources,
the number of fields is increased by 43 thanks to the 2MASS
coverage of the northern ISOGAL fields. On the other hand, the
total numbers of associated sources are comparable, which means
that there are fewer sources on average in each field. This
decrease in average density of sources in the ISOGAL-2MASS catalog
can be attributed to the stricter constraint on photometric
quality for the 2MASS PSC sources. This ISOGAL-2MASS PSC catalog
is available via the VizieR Service at the Centre de Donn\'ees
Astronomiques de Strasbourg (CDS,
http://vizier.u-strasbg.fr/viz-bin/VizieR/).

\section{Method}

\subsection{Assumptions}

We use the method described in detail by \citet{Jiang03} to
derive the extinction values.  In the following sections, we adopt the following
symbol convention: magnitudes in the J,Ks, bands and at 7 and
15${\rm \mu}$m are denoted by J, Ks, [7] and [15] respectively;
intrinsic color indexes {\rm
$m_{\lambda_{1}}^{0}-m_{\lambda_{2}}^{0}$ } by
C$_{\lambda_{1}\lambda_{2}}^{0}$; observed color indexes by
C$_{\lambda_{1}\lambda_{2}}$; the color excess of the observed to the
intrinsic color {\rm
C$_{\lambda_{1}\lambda_{2}}^{0}-C_{\lambda_{1}\lambda_{2}}$} by
E($\lambda_{1}-\lambda_{2}$); the extinction by A$_{\lambda}$. The method is
based on two assumptions. The first assumption is that the group comprising
RGB-tip and early AGB stars have approximately constant intrinsic color
indexes C$_{\rm JKs}^{0}$, C$_{\rm Ks7}^{0}$, C$_{\rm Ks15}^{0}$. The second
is that the ratio of color excess E(Ks-[7]) to E(J-Ks),i.e.
k$_{7}\equiv$E(Ks-[7])/E(J-Ks), is constant in one field. Then the
extinction at 7${\rm \mu}$m can be calculated by a linear fitting of the
observed color indexes C$_{\rm Ks7}$ to C$_{\rm JKs}$, i.e.
\begin{equation}
C_{\rm Ks7}-C_{\rm Ks7}^{0} =  k_{7}\times(C_{\rm JKs}-C_{\rm
JKs}^{0})
\end{equation}
\begin{equation}
A_{\rm Ks}-A_{7}=k_{7}\times(A_{\rm J}-A_{\rm Ks})
\end{equation}
The same formula can be applied to the 15${\rm \mu}$m case.

\begin{figure}
  \includegraphics[bb=0 0 256 256, width=10cm]{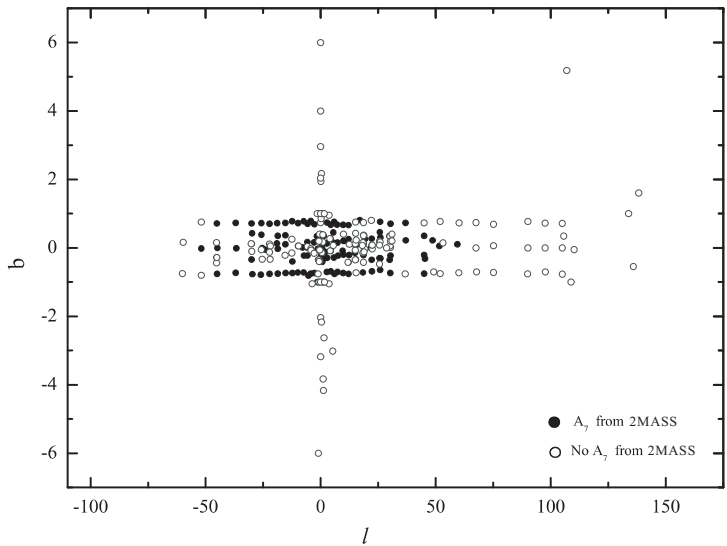}

  \caption{Distribution of the ISOGAL fields in the Galactic plane. The
  filled circles denote the fields for which reliable results on the 7$\mu$m
  extinction were obtained and the open circles denote the fields for which
  no reliable results on the 7$\mu$m extinction could be obtained.}

  \label{fig1}
\end{figure}

\subsection{Selection of sources}

Two types of sources are to be excluded because their intrinsic colors
differ from our standard red giant sources: i) dusty young stellar
objects and AGB stars with thick circumstellar envelopes, and ii) foreground
early-type stars. The former are distinguished by their color index
C$_{715}^{0}$ which is usually larger than 0.4. Thanks to the small
extinction at both 7 and 15${\rm \mu}$m, the observed color index C$_{715}$
may be used as a good approximation to the intrinsic one C$_{715}^{0}$.
\citet{Felli02} calculated the influence of the circumstellar matter on
C$_{715}^{0}$ and found that the sources having this color index greater
than 0.4 may have thick circumstellar matter and therefore redder intrinsic
C$_{\rm Ks15}^{0}$ and C$_{\rm Ks7}^{0}$. However, selection by color is
applicable only to the sources in the ``FC'' fields which were observed at
both 7${\rm \mu}$m and 15${\rm \mu}$m. It is not suited to the sources in
the ``FA'' or ``FB'' fields observed at only one wavelength since no color
index C$_{715}$ can be derived. Most foreground early-type stars can be
excluded simply by dropping the sources with C$_{\rm JKs}<$1.2.

Some fields lie in directions with small extinction, as shown by their small
E$({\rm J-K_{S}}$) values. From the analysis of the extinction law at 7${\rm
\mu}$m and 15${\rm \mu}$m by
\citet{Jiang03}, A$_{7}$ is about 0.18E$({\rm J-K_{S}})$ and
A$_{15}$ is about 0.16E$({\rm J-K_{S}})$. Thus, an extinction of the order
of E$({\rm J-K_{S}}$)= 1 mag results in A$_{7} \sim$ 0.18 mag, comparable to
the photometric error. In order to be able to obtain significant extinction
in these mid-infrared bands, only the fields with much greater near-infrared
extinctions are targeted. Only sources with large values of the color excess
E$({\rm J-K_{S}})$ are chosen so that their extinction at mid-infrared
wavelengths will be significant. This will be further clarified in the next
section.

In summary, a source to be selected should have: good photometry quality
both in the ISOGAL mid-infrared and DENIS or 2MASS near-infrared bands, good
near-infrared-mid-infrared association flags, and a moderate mass loss rate
with C$_{715}<$0.4. In addition, for a field to be qualified, there must be
enough sources which suffer significant and detectable extinction in the
mid-infrared bands.

\begin{figure}[ht]
\includegraphics[width=17cm]{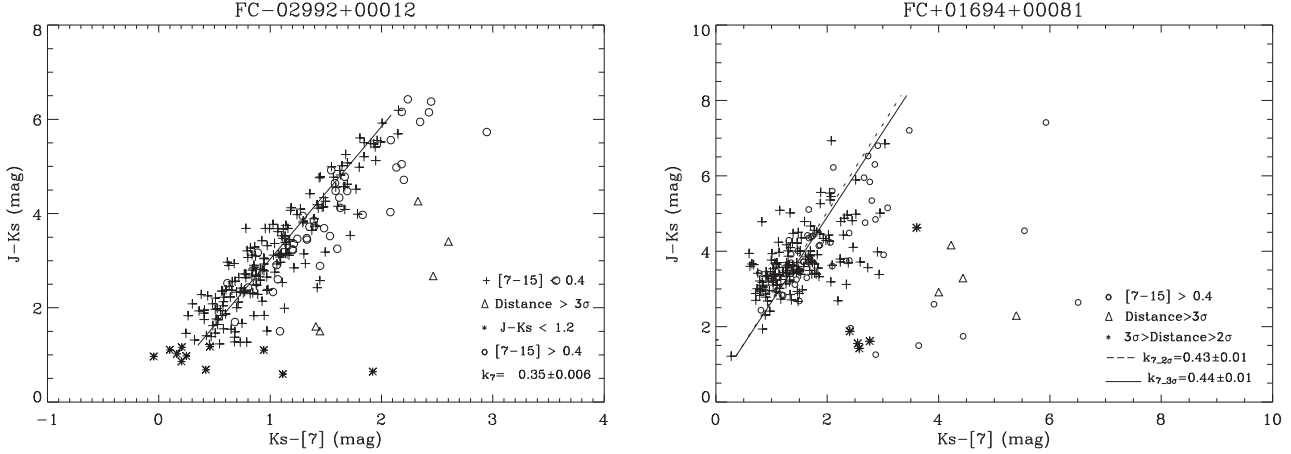}

  \caption{Two example fields showing all the ISOGAL-2MASS associated sources.
  The crosses denote the sources kept for final fitting. The other symbols
  denote various kinds of sources rejected before fitting, specifically, the
  asterisks for J-K$_{\rm S}<$1.2, open circles for [7]-[15]$>$0.4, and
  triangles for deviations from the fitted line greater than 3$\sigma$.  The
  figure on the left shows a good-quality case, FC-02992+00012, with small
  $\sigma_{\rm k_{7}}$; the figure on the right shows a
  mediocre case, FC+01694+00081, with relatively large error in k$_{\rm 7}$.
  In the right figure, the points with deviation between 2$\sigma$ and
  3$\sigma$ are shown also. The results from fitting by the 2$\sigma$ and
  3$\sigma$ rejections (denoted by dash and solid lines respectively) are
  very close to each other (see \ref{prfit}). }

  \label{fig2}
\end{figure}

\subsection{Fitting Procedure}
\label{prfit}

Fitting is done in two steps. First, we make a preliminary robust
fitting to all the points which satisfied the previous criteria.
In comparison with normal least-square linear fitting, the robust
fitting method, which minimizes the absolute deviation, is less
sensitive to small number of large outliers and more appropriate
for our cases which usually have fewer sources with large color
indexes. This fitting distinguishes the sources with large
deviation, i.e. deviation larger than 3$\sigma$ from the fitted
line. These sources were rejected because they have little
probability of obeying the linear relation from the statistical
point of view; on the other hand, most of them have relatively
large C$_{\rm Ks7}$, which indicates the likelihood of extinction
arising from circumstellar dust envelopes. The linear fitting was
carried out both ways, one fitting C$_{\rm Ks7}$ (as y) to C$_{\rm
JKs}$ (as x) and the other fitting C$_{\rm JKs}$ (as y) to C$_{\rm
Ks7}$ (as x); a source deviating from either fitting larger than
3$\sigma$ was rejected. We also tested 2$\sigma$ as the critical
deviation for rejection and found the results were little changed.
As an example, the right panel in Fig.~\ref{fig2} shows the
sources with deviation between 2$\sigma$ and 3$\sigma$ as well as
the results by fitting sources with both criteria. It can be seen
that the results are very close to each other.

After the rejection of the sources with large deviations, 55 of
the 7$\mu$m fields (and 94 of the 15$\mu$m fields) were discarded
using the following two criteria in combination with the preceding
selection of sources: a field must have (1) at least 30 sources
remaining and (2) values of E$_{\rm JKs}$ spanning more than 2
magnitudes, i.e. the maximum C$_{\rm JKs}$ should be larger than
at least 3.2, since the minimum of C$_{\rm JKs}$ is set at 1.2 mag
which agrees with the choice of RGB tip or early AGB stars (see
succeeding paragraphs for explanation). The first condition is
required for the results to be statistically meaningful. The
second condition means that the extinction at 7 and 15${\rm \mu}$m
is large enough to be detectable. If we adopt our previous result
that A$_{7}\sim 0.2{\rm E}_{\rm JKs}$, then E$_{\rm JKs}$ must be
$>2$, i.e. A$_{V}>12$mag, and A$_{7}>0.4$, just about twice the
ISOGAL photometry rms accuracy at 7${\rm \mu}$m.

These conditions retain about 73\% of the 7$\mu$m fields and 57\%
of the 15$\mu$m fields. The resulting set of data covers 151
fields (including 35 FA and 116 FC fields) suitable for the 7${\rm
\mu}$m determination  and 126 fields (including 34 FB and 82 FC
fields) for the 15${\rm \mu}$m determination, based on the
ISOGAL-2MASS PSC. For the ISOGAL-DENIS PSC, the retained data
cover 139 fields (36 FA plus 103 FC fields) with 7${\rm \mu}$m
photometry, and 116 fields (42 FB plus 74 FC fields) with 15${\rm
\mu}$m photometry.

Two examples showing rejected sources are shown in Fig.~\ref{fig2}. In this
figure, the left panel shows the sources from the field FC-02992+00012,
which represents a good case with small error in k$_{\rm 7}$; the right
panel shows the sources from the field FC+01694+00081, which represents a
mediocre case with relatively large error in k$_{\rm 7}$.

In principle, both the slope and the intercept, thus the color
excess ratio, can be derived from the fitting. However, we choose
to fix the intrinsic color index C$_{\rm Ks7}^{0}$ or C$_{\rm
Ks15}^{0}$. There are two reasons for doing this. First, the
intrinsic color indexes C$_{\rm Ks7}^{0}$ and C$_{\rm Ks15}^{0}$
of RGB tip stars or early AGB stars have small scatter, with a
dispersion of about 0.2mag comparable to the photometric
uncertainty \citep{Jiang03}. Second, for many fields, when C$_{\rm
Ks7}^{0}$ and C$_{\rm Ks15}^{0}$ are not fixed, the fitting leads
to unrealistic values of C$_{\rm Ks7}^{0}$ and C$_{\rm Ks7}^{0}$
because of the lack of good quality low extinction sources. The
adopted values of C$_{\rm Ks7}^{0}$ and C$_{\rm Ks15}^{0}$ are
0.35 and 0.40 respectively. The choice is based on the calculation
of a 4000~K blackbody radiation, which gives a value in good
agreement with the observations in the ISOGAL field
FC$-$01863+00035 with particularly good quality data
\citep{Jiang03}.  In addition, C$_{\rm JKs}^{0}$ is also fixed to
be 1.2 mag, since the observations to the fields in Baade's window
by \citet{Glass99} and the modelling by \citet{Bertelli94} both
showed that this index of such late-M type stars is 1.2$\pm$0.1.
Therefore, the fitted line is forced to pass through the point
[0.35,1.2] or [0.40,1.2] in the K$_{\rm S}$-[7] vs. J-K$_{\rm S}$
or K$_{\rm S}$-[15] vs. J-K$_{\rm S}$ diagram, respectively. Only
the slope of the fitted line, k$_{7}$ or k$_{15}$, is derived
during the fitting.  Even with the previously described
restrictions, the fitting results still appeared poor in some
cases, which may be attributed to some restriction not being
appropriate. For example, C$_{\rm Ks7}^{0}$ is fixed, which is not
really true, since C$_{\rm Ks7}^{0}$ is likely to be slightly
different from case to case. The last selection is based only on
fitting error, without relation to any physical meaning. The final
sample requires a relative error of k$_{\rm 7}$ less than 5
percent, i.e. {\rm $\sigma_{\rm k_7}/{\rm k_7} < 0.05$}, which
takes into account the error induced in the fitting procedure.
This restriction excluded 22 fields. For the 15$\mu$m case, the
relative error of k$_{\rm 15}$ is constrained to be $<$10\% in
order to retain sufficient data, as 6 fields are excluded. Thus
results with small fitting errors are derived for 129 fields and
120 fields, around 7 and 15$\mu$m respectively.

From k$_{7}$, the ratio of the color excess E(Ks-[7]) to E(J-Ks)
is derived. To obtain the extinction A$_{7}$, one more step is
needed as can be seen from equation (2) in Section 3.1. The value
of A$_{7}$ depends not only on the slope k$_{7}$, but also on the
near-infrared extinction values, A$_{\rm J}$ and A$_{\rm Ks}$. The
influence of near-infrared extinction values was discussed in
\citet{Jiang03}. Generally, the extinction law in the
near-infrared is accepted as universal with no variation from
sightline to sightline \citep{Draine89}. Nevertheless, the
measurements of the near-infrared extinctions differ a little from
group to group. The values given by \citet{Glass99} and
\citet{Hulst46} are in good agreement, while the values given by
\citet{Rieke85} (R\&L) are slightly higher. \citet{Jiang03}
adopted the former while \citet{Lutz99} and \citet{Indebetouw04}
adopted the later. The average values of A$_{7}$ expressed in
A$_{\rm Ks}$ are 0.34 when using the values \ by \citet{Glass99}
and \citet{Hulst46}, while being 0.46 when using the values by
\citet{Rieke85}. For reason of consistency, we would prefer to use
the Glass figures. However, this group of values give rise to some
negative A$_{7}$ (7${\rm \mu}$m) values, which is obviously
unreasonable, and may mean the Glass values underestimate the
near-infrared extinctions. So instead of the previous choice, made
by by \citet{Jiang03}, the near-infrared values by R\&L are used.
Although the extinctions at 7${\rm \mu}$m relative to A$_{V}$
depend somewhat on the near-infrared extinctions, the value
normalized to the extinction in the Ks band is less dependent on
it. Therefore, we express the extinction at 7${\rm \mu}$m in terms
of the extinction in the Ks band. In Table \ref{exttab}, the
results from different near-infrared extinction values are shown
in detail.

\begin{table}[h]
\caption[]{Average extinction values in the LW2~(7{\rm $\mu$m})
and LW3~(15{\rm $\mu$m}) bands based on three different
near-infrared extinctions }
\begin{tabular}{lccc} \hline
Band & \multicolumn{3}{c}{\rm $A_{\lambda}/A_{Ks}$} \\ &
 vdH\footnotemark[1]\ & ISG\footnotemark[2] & R \& L\footnotemark[3] \\
 \hline J & 2.816 & 2.876 & 2.509 \\
      K$_{\rm S}$ & 1 & 1 & 1 \\
      7 & 0.37 & 0.35 & 0.47 \\
     15\footnotemark[4] & 0.30 & 0.28 & 0.40\\ \hline
\end{tabular}

\vspace*{1mm}
\begin{small}
\noindent \footnotemark[1]Values for J and K$_{\rm S}$ derived
from van de Hulst (1946) curve\\
\footnotemark[2] Values for J and K$_{\rm S}$ derived from
Glass (1999) (update of van de Hulst's values)\\
\footnotemark[3] Values for J and K$_{\rm S}$ derived
by Rieke \& Lebofsky (1989) for stars towards the Galactic center\\
\footnotemark[4] The average is derived only from the LW3 fields.
See Section 4.3 for reference.
\end{small}
\label{exttab}
\end{table}

\section{Results and Discussions}

\subsection{Extinctions at 7${\rm \mu}$m}

For the 115 of all the selected fields observed by both DENIS and
2MASS, we have attested that the values derived for A$_{7}/A_{\rm
Ks}$ from the two sets of near-infrared data are consistent. They
display the same average value with an rms difference equal to
0.04. Only two or three fields displayed anomalous differences for
A$_{7}/A_{\rm Ks}$ and these arise from a problem with the DENIS
data. Thus, the two sets of data yield practically identical
results. However, since 2MASS completely covers the ISOGAL-DENIS
fields and includes some northern ISOGAL fields not included by
DENIS, the results derived from the ISOGAL-2MASS PSC represent the
entire sample and only they are considered hereafter.

From the above choices of the intrinsic color indexes C$_{\rm
Ks7}^{0}$ and C$_{\rm Ks15}^{0}$ as having the values 0.35 and
0.40 respectively, C$_{715}^{0}$ is inferred to be 0.05. This
corresponds to the case of stars with very little dust. However,
the criterion C$_{715}<0.4$ was set to exclude the effect of
circumstellar dust on the brightness at 7$\mu$m and 15$\mu$m, on
the color indexes C$_{\rm Ks7}$ or C$_{\rm Ks15}$, and thus on the
result. The critical value of  C$_{715}$ affects the A$_{\rm 7}$
result only slightly, while A$_{\rm 15}$ changes a little more.
When the critical value is changed from the present 0.4 to 0.6,
the average of A$_{\rm 7}$ is decreased by only 0.01, and the
average of A$_{\rm 15}$ is decreased by 0.08; when it is changed
from 0.4 to 0.2, the average of A$_{\rm 7}$ is increased by 0.004,
and the average of A$_{\rm 15}$ is increased by 0.09. Such
tendencies are expected: radiation by circumstellar dust increases
the 7$\mu$m and particularly the 15$\mu$m fluxes, which imitates a
decrease in the corresponding extinctions. Thus, including more of
the sources with circumstellar dust envelopes would underestimate
the extinction. As the result on A$_{\rm 7}$ is little affected,
the value chosen appears to be small enough to exclude the effect
of dusty envelopes. From the fact that A$_{\rm 15}$ is affected
when the critical value is changed, it can be seen that this
criterion cannot completely remove the influence of dust shells.
However, if this critical value is moved towards the blue, the
number of sources and fields would decrease significantly and thus
lower the statistical significance of the results. Therefore, the
value of 0.4 is kept for both bands accordingly.

The extinction values A$_7$ are displayed in Table \ref{tabext7} [the full
table is given for the selected 129 fields in the electronic version at CDS
(http://vizier.u-strasbg.fr/viz-bin/VizieR]. From left to right, the table
includes the ISOGAL field name, filter (``2'' stands for ``LW2'' and so on),
the number of sources included in the final fitting, the slope k$_{\rm 7}$,
the 1-sigma error of k$_{\rm 7}$, and the extinction value at 7$\mu$m
relative to A{\rm $_{\rm K_S}$}. There are 106, 16 and 7 fields with the
filters LW2, LW6, and LW5, respectively.

\begin{figure*}
  \includegraphics[width=16cm]{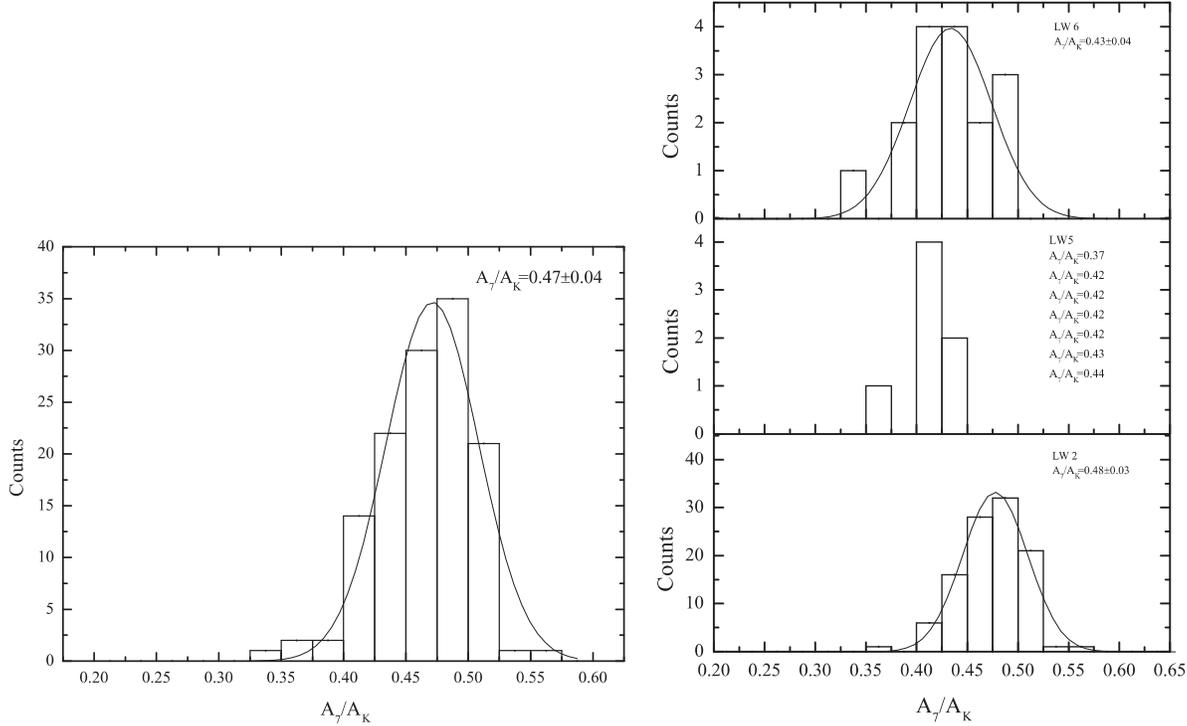}
  \caption{Distribution of all the A$_{7}$ values as a whole (left panel)
  and in the different filters (right panel).}
  \label{fig4}
\end{figure*}

From Table \ref{tabext7} and Fig. \ref{fig4}, it can be seen that
 A$_{\rm 7}$/A$_{\rm Ks}$ has a distribution which can be well
approximated by a Gaussian function, with the central value of
0.47A$_{\rm Ks}$ and a FWHM of about 0.07A$_{\rm Ks}$. The minimum
and maximum are 0.33A$_{\rm Ks}$ to 0.55A$_{\rm Ks}$ respectively.
Such variation includes the differences in filter wavelength
ranges, statistical ``noise'' in determining k$_{\rm 7}$,
systematic errors and the possible variation of A$_{\rm
7}$/A$_{\rm Ks}$ along various lines of sight. We will try to
disentangle these various contributions in the two following
sections.

\subsubsection{The central value of A$_{\rm 7}$}

The average value of the distribution of A$_{\rm 7}$ (c.f.
Fig.\ref{fig4}) we refer to as the central value is 0.47A$_{\rm
Ks}$. This value is very consistent with that of \citet{Lutz96} --
an extinction of 0.45A$_{\rm Ks}$ at 6.8$\mu$m. It also agrees
well with the value 0.48A$_{\rm Ks}$ at both 5.7$\mu$m and
7.8$\mu$m, derived for the $l=42\deg$ sightline using the GLIMPSE
data by \citet{Indebetouw04}.

\begin{figure*}
  \includegraphics[width=14cm]{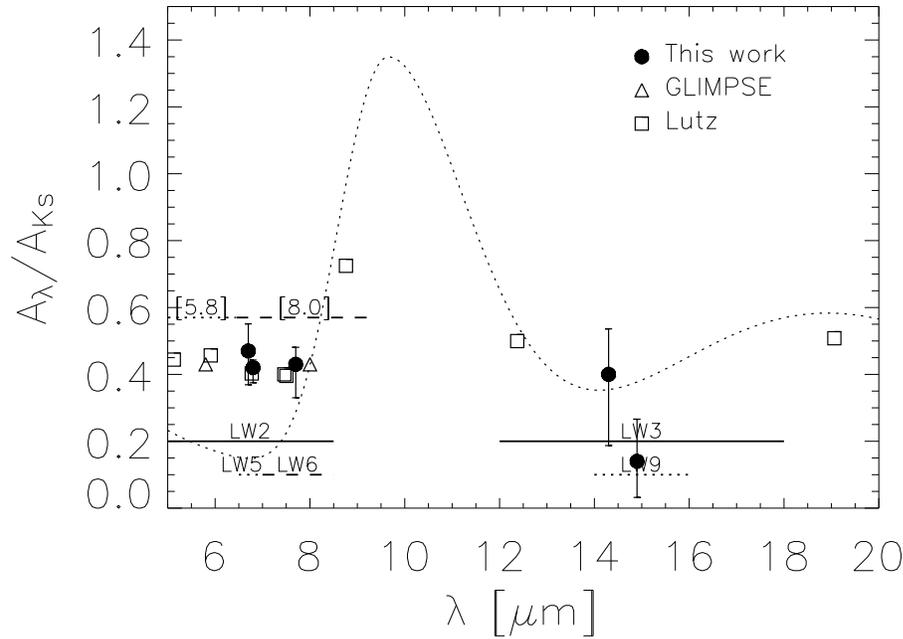}

  \caption{Comparison of this work (filled dots)with the results of
  \citet{Lutz99}(open squares) and GLIMPSE by \citet{Indebetouw04}(open
  triangles). The values of GLIMPSE that are shown are the average ones. The
  horizontal lines denote the spectral coverages of the ISOCAM and
  Spitzer IRAC filters. The vertical bars show the ranges of the values of
  A$_{7}$ and A$_{15}$.  The dotted curve displays the silicate features
  calculated from Rosenthal's analytic formula. }

  \label{comp}
\end{figure*}

As we mentioned in the Introduction, the ISOGAL observation used
three different filters centered at about 7${\rm \mu}$m, viz.\
LW2, LW5 and LW6. To see if the extinction around 7$\mu$m varies
with wavelength, we have first determined the mean values of
extinction in the three bands. They are 0.48, 0.42 and 0.43A$_{\rm
Ks}$ for LW2, LW5 and LW6, respectively.  The standard deviations
of A$_{7}$ in these three cases are 0.016, 0.020 and 0.019,
significantly smaller than the variation ranges of A$_{7}$ derived
for the corresponding filters. The results are plotted in
Fig.~\ref{comp}, and compared with the values based on
Spitzer/GLIMPSE data by \citet{Indebetouw04} and ISO data by
\citet{Lutz99}. It is seen that our results are in good agreement
with others.

\subsubsection{Variation of A$_{\rm 7}$}

In order to discuss whether the spread in the values of A$_{\rm 7}$/A$_{\rm
Ks}$ exceeds that due to the data noise, and whether it provides evidence
for extinction variations along differing lines of sight, we first focus on
the LW2 data which are by far the most complete and homogeneous, with the
best quality. The filters LW6 and LW5 were used only in the most difficult
and crowded fields to reduce the effects of saturation by strong sources
which are particularly numerous in star forming regions. The distribution of
the values obtained with LW2 for A$_{\rm 7}$/A$_{\rm Ks}$ is reasonably well
fitted by a Gaussian distribution with a central value of 0.48 and an r.m.s.
of 0.03 (see Fig.\ref{fig4}).

Table \ref{tabext7} gives the value of the r.m.s. $\sigma_{\rm
k_{7}}$ of the slope k$_{\rm 7}$ from the linear fitting of the
relative reddening J-Ks/Ks-[7] (Fig. \ref{fig2}). The r.m.s.
values include mainly the noise appropriate to the various
magnitudes, but also other kinds of variations such as uncertainty
in the intrinsic colors, source variability, extinction variation
along a given line of sight, etc.  For all these 7$\mu$m fields,
the average of $\sigma_{k7}$ is 0.011, while k$_{7}$ ranges from
0.30 to 0.44, with an unweighted average of 0.355 (weighted
average 0.352) and standard deviation 0.025.  Such a value (0.025)
appears significantly, albeit somewhat marginally, larger than the
r.m.s. $\sim$0.011 of k$_{\rm 7}$. There is thus some evidence
that A$_{\rm 7}$/A$_{\rm Ks}$ may vary significantly, depending on
the line of sight.

The range [0.37A$_{\rm Ks}$, 0.55A$_{\rm Ks}$] (or even
0.33A$_{\rm Ks}$ if one includes LW6 fields) of A$_{\rm 7}$ from
our analysis covers that obtained by \citet{Indebetouw04} from
0.37 at the $l=284\deg$ sightline to 0.48 at the $l=42\deg$
sightline at 7.8$\mu$m. The value of 0.40A$_{\rm Ks}$ around
7$\mu$m from \citet{Lutz99} is also inside this range.

It has been known for long time that the selective extinction in
the visual and ultraviolet range varies from one sightline to
another, depending on the interstellar environment through which
the starlight passes. The representative parameter {\rm R$_\equiv
A_{V}/(A_{B}-A_{V})$ } changes from values as low as 2.1 (toward
very diffuse interstellar clouds) to values as large as 5.6-5.8
(toward very dense interstellar clouds). Such variation has been
explained by different size distributions among the grains, i.e.
the larger R$_{V}$ distribution has significantly fewer small
grains than the smaller R$_{V}$ distribution, which is expected
since generally there is relatively less extinction at short
wavelengths (provided by small grains) for larger values of
R$_{V}$.

However, in the near-infrared, the available observations are consistent
with a uniform extinction law at wavelengths shortward of 3.5${\rm \mu}$m.
\citet{Cardelli89} concluded that the shape of extinction law for 0.7${\rm
\mu m<\lambda<3.5\mu m}$ was independent of R$_{\rm V}$ within the errors of
measurements. It is further inferred from the constancy of the near-infrared
extinction law that the size distribution of the largest particles are
almost the same in all directions \citep{Mathis90}. Following this opinion,
one would initially expect no variation of the extinction law in
mid-infrared, since the mid-infrared extinction must be attributed to larger
grains. However, with the grain models which have a size distribution of
log-normal form \citep{Weingartner01} rather than classical power law form,
\citet{Draine03} (Figure 4) showed that the extinction law in
mid-infrared could be much more dependent on R$_{\rm V}$ even though the law
in near-infrared is rather independent of R$_{\rm V}$. Similar to the
situation at UV and visual wavelengths, the larger R$_{\rm V}$ value results
in larger extinction in the 2.5-8${\rm \mu}$m range. At the wavelength
7${\rm \mu}$m of our interest, the extinction changes by a factor of about
two when R$_{\rm V}$ varies from 3.1 to 5.5. Thus in principle, the
variation of the selective extinction around 7${\rm \mu}$m is
understandable, considering the variation in the dust grain size
distribution. On the observational side, \citet{Indebetouw04} found slight
but significant differences in the wavelength dependence of the mid-infrared
extinction from two regions. One possible explanation for this discrepancy
between the near-infrared constancy and the mid-infrared variation of the
extinction law is that the extinction at 7${\rm \mu}$m is already affected
by the silicate 9.7${\rm \mu}$m feature. While the continuum extinction
(attributed to the graphite grains, which are the only source of
extinction in the near-infrared wavelength range) does not vary from
sightline to sightline, the silicate feature extinction may change in
different environments. It is well accepted that there is variation in
A$_{V}/\Delta\tau_{9.7}$. Studies of the 9.7${\rm
\mu}$m feature yielded A$_{V}/\Delta\tau_{9.7}$ = 9 towards the Galactic
center direction \citep{Roche85} and about 18.5 toward WC stars (see e.g.
\citet{Hucht96}). In addition, within the Taurus molecular cloud, this ratio
varies substantially \citep{Whittet88}. Such variations in strength of the
9.7${\rm \mu}$m feature and its profile would surely affect the selective
extinction in mid-infrared. Nevertheless, the wavelength limit of our 7${\rm
\mu}$m filters remains less than 8.5$\mu$m and is thus only at the edge of
the silicate 9.7${\rm \mu}$m feature, where the relative extinction caused
by silicate to the peak (A$_{\lambda}$/A$_{\rm peak}$) is below
0.1\citep{Draine89}. The influence of the silicate dust on the 7${\rm \mu}$m
extinction should therefore not be very significant. This may be the reason
why the variation of the 7${\rm \mu}$m extinction is not very large.

In order to look for further evidence that the variation of the 7${\rm
\mu}$m extinction law is caused by variable silicate extinction, one could
try to compare the extinctions deduced for the different filters used in the
ISOGAL observation. Because the silicate feature centers at 9.7${\rm \mu}$m
and starts from about 7.5${\rm \mu}$m, a small part of the LW2 filter and a
major part of the LW6 filter's bandwidths overlap the silicate feature,
while the narrow-band filter LW5 is completely outside it.  In the right
panel of Fig. \ref{fig4}, the distributions of A$_{7}$ in three filters are
shown.

However, it is not certain that such a comparison is relevant because the
number of LW6 and LW5 observations is relatively small, with data of poorer
quality, and they cover atypical regions of the sky such as the Galactic
Center neighborhood and star forming regions. There is no obvious difference
between the LW5 and LW6 values; however, the number of LW5 values is too
small to infer any real conclusion. The values of A$_{7}$ from the LW2 and
LW6 data both have approximately Gaussian distributions with similar widths.
The most distant point in the A$_{7}$ distribution of the LW6 observations
is from the field FC+01694+00081. From Fig.~\ref{fig2} it can be seen that
the fitting for the extinction is of relatively poor quality. Indeed this
field is located in the direction of the extremely young open cluster
NGC~6611 \citep{Belikov99}, which may be the reason for the large scatter in
the J-Ks/Ks-[7] plot and thus the poor quality of the fit. The difference
between the LW2 and LW6 groups is apparent in the central values of the
distributions. The LW6 group values are centered at 0.43A$_{\rm Ks}$ and the
LW2 group values center at 0.48A$_{\rm Ks}$, i.e. the extinction in the LW2
band is slightly larger than that in the LW6 band. However, it seems very
difficult when explaining the origin of such a difference, if real, to
disentangle a possible wavelength effect from the fact that the LW6
observations are towards very special regions.

\subsection{Correlation with locations in the Galactic plane}

\begin{figure*}
\includegraphics[bb=0 0 560 226, width=16cm]{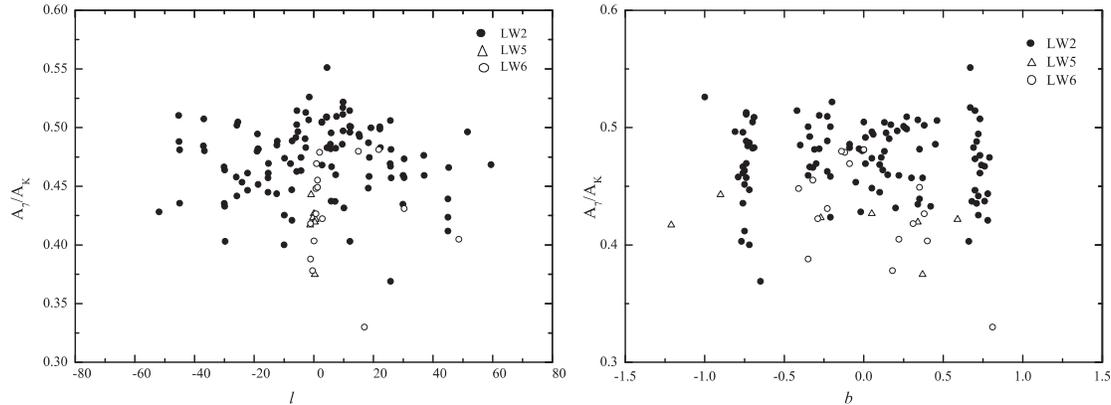}

  \caption{Distribution of A$_{7}$ with Galactic longitude (left panel)
  and latitude (right panel). Different symbols denote different filters:
  dot for LW2, triangle for LW5 and circle for LW6. }

  \label{fig6}
\end{figure*}

Shown in Fig. \ref{fig6} are the distributions of A$_{7}$ with Galactic
latitude and longitude. Because the ISOGAL project targeted fields in the
Galactic plane, the latitude range is small, and the distance from the
midplane is always less than 1\degr. There is no systematic change of
A$_{7}$ with latitude seen in this latitude range.

The longitude range of the fields we have retained covers about
$-$60\degr to +60\degr. There is no clear evidence that the
extinction changes systematically with longitude within this large
range. The measurements in the bulge direction are abundant and
the results look rather scattered. The random variations of
A$_{7}$ may arise partly from ``noisy'' data. However, they are
not incompatible with moderate local variations of the extinction
caused by patchy distribution of the interstellar medium on small
scales.

\subsection{Extinctions at 15${\rm \mu}$m}

Different from the 7${\rm \mu}$m extinction, the 15${\rm \mu}$m extinction
is probably dominated by the combined effect of spectral features of
silicate dust with a minor contribution from continuum extinction. If the
continuum extinction followed the power law A$_{\lambda}\propto
\lambda^{-1.7}$, it would amount to 0.04A$_{\rm Ks}$ at 15${\rm \mu}$m.
However, the silicate spectral features can bring about much larger
extinction. For example, according to the analytic formula of
\citet{Rosenthal00}, the extinctions at 15$\mu$m caused by the 9.7${\rm
\mu}$m and 18${\rm \mu}$m features are {\rm $\sim0.04A_{K}$} and {\rm
$\sim0.30A_{K}$} respectively. If further the spectral response of
the ISOCAM filter is taken into account, the convolution of the spectral
transmission profile of the LW3 and LW9 filters with Rosenthal's formula
yields 0.44A$_{\rm Ks}$ and 0.39A$_{\rm Ks}$ respectively. Thus the 15${\rm
\mu}$m extinction is mainly contributed by the silicate spectral feature
extinction, with the 18${\rm \mu}$m feature being the major contributor. In
addition, because of the variation of the silicate features in different
environments, the 15${\rm \mu}$m extinction law is expected to vary as well.

Because the flux at 15${\rm \mu}$m is more affected by circumstellar dust
than at 7${\rm \mu}$m, excluding the sources with relatively thick
circumstellar envelopes becomes more important. For this reason, only the FC
fields observed at both 7 and 15${\rm \mu}$m with [7]-[15]$<$0.4 are chosen.
Using this constraint, most of the sources with thick circumstellar
envelopes are expected to be rejected. We adopted an intrinsic color index
C$_{\rm Ks15}^{0}$ of 0.40, because a 4000~K (typical for a red giant star)
blackbody radiation has this value and it is also the result deduced from
the previously studied ISOGAL field with good quality data by
\citet{Jiang03}.

\begin{figure}
  \includegraphics[width=14cm]{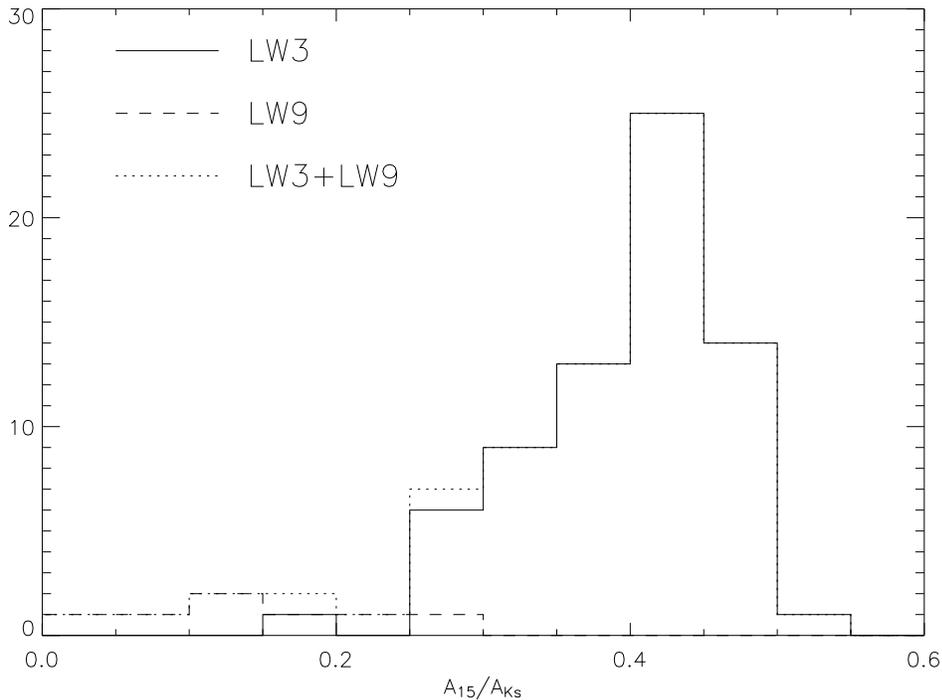}

  \caption{The histogram of the selective extinction at 15${\rm \mu}$m. The solid line shows
  the results from the fields observed in the LW3 filter, the long-dash line
  the LW9 filter, and the dotted line the sum of the two cases. }

  \label{fig7}
\end{figure}

The final results on the 15${\rm \mu}$m extinction derived from
the FC fields are shown in Table \ref{tabext15} [the full table is
given for the selected 76 fields in the electronic version at CDS
(http://vizier.u-strasbg.fr/viz-bin/VizieR] and displayed as a
histogram in Figure \ref{fig7}. Among the 76 FC fields for
measuring A$_{\rm 15}$ that remain, 69 were observed with the LW3
filter and 7 with the LW9 filter. For the 69 LW3 fields (i.e.
observed with the LW3 filter), the average value of A$_{\rm 15}$
is 0.40~A$_{\rm Ks}$ with a standard deviation of 0.07. This is
remarkably consistent with the expected values from the
superposition of the continuum extinction and the silicate
spectral extinction as analyzed in the beginning of this section.
This result is also in agreement with the value 0.41A$_{\rm Ks}$
from \citet{Jiang03} for one line of sight. As $\sigma_{\rm A_{\rm
15}}$=0.037, and A$_{\rm 15}$ ranges from 0.19A$_{\rm Ks}$ to
0.54A$_{\rm Ks}$, which is larger than 3$\sigma$, it is possible
to attribute part of the range to the field-to-field differences.
This is consistent with the expectation that the silicate feature
varies with the interstellar environment. However, the
distribution of A$_{\rm 15}$ is asymmetrical (see Figure
\ref{fig7}); most of the results are within 3$\sigma$. Taking into
account the much larger rms of A$_{\rm 15}$, the reality of the
variation of A$_{\rm 15}$ must be proved more rigorously. For the
LW9 fields, the small number of fields makes the statistics less
meaningful. Nevertheless, the average value of A$_{\rm 15}$ for
them is 0.14A$_{\rm Ks}$ with a standard deviation of 0.09. Shown
in Fig. \ref{fig7} are the distributions of A$_{\rm 15}$ for the
LW3, LW9 fields separately and together. As can be seen in Fig.
\ref{comp}, the value for the LW9 fields is clearly smaller than
for the LW3 fields and significantly smaller than the theoretical
expectation.  The difference in A$_{\rm 15}$ between LW9 and LW3,
as well as the difference between LW9 and the value from
Rosenthal's analytic formula, is larger than the uncertainty of
the fitting error. It may be concluded from the former that the
extinction decreases from 14.3$\mu$m (the reference wavelength of
LW3) to 14.9$\mu$m (the reference wavelength of LW9) and, from the
latter, that Rosenthal's expression overestimates the extinction
around this waveband. But we should be cautious about the
extinction result at LW9, because: (1) The LW9 fields are
relatively peculiar regions with generally strong star forming
activity so that the extinction may differ from diffuse
interstellar regions. As Rosenthal's analysis is towards an area
with an A$_{\rm Ks}$ of only one magnitude, the difference may be
attributable to the different environment. Further observations of
some diffuse regions are needed. (2) The radiation from the
circumstellar dust is strong around 15$\mu$m, and its effect can
not be purged from our selection of the color index, as discussed
in Section 4.1.

\section{Summary}

The ISOGAL project surveyed about 20 square degrees in the Galactic plane
and detected numerous evolved stars with similar intrinsic infrared colors.
Based on the point sources in the ISOGAL PSC catalog, we have built a
cross-identified catalog between the ISOGAL and the 2MASS PSCs similar to
the ISOGAL-DENIS catalog of Schuller et al. (2003). It includes about 60,000
sources. After selection from the mid-infrared and near-infrared colors and
photometry quality flags, the extinctions around 7 and 15${\rm \mu}$m were
derived towards about 129 fields. The central value of the 7${\rm \mu}$m
extinction is about 0.47A$_{\rm Ks}$ with a width of about 0.07A$_{\rm
K_{S}}$ (using the near-infrared extinctions of \citet{Rieke85}). The
average extinction shows some variation in the small range of wavelength
defined by the LW2, LW5 and LW6 filters, with the value from the LW2 filter
slightly larger than from the LW5 and LW6 filters. In addition, there is a
possibility that A$_{\rm 7}$ varies along different lines of sight, falling
within the range of [0.33, 0.55]A$_{\rm Ks}$, with the mean fitting error
for any particular field being about 0.02.  The values from H recombination
lines toward the Galactic Center found by \citet{Lutz96} fall within the
range of our results, as does the recent result from the Spitzer GLIMPSE
data at $l=42\deg$ \citep{Indebetouw04}.

The extinction at 15${\rm \mu}$m was found to be about 0.4A$_{\rm Ks}$ based
on the data from fields observed both at 7 and 15$\mu$m. Although this
result suffers relatively large uncertainty due to scatter in the intrinsic
color C$_{\rm Ks15}^{0}$, it is in reasonable agreement with theoretical
expectation.

\begin{acknowledgements}
This paper made use of data products from the Two Micron All Sky
Survey, which is a joint project of the University of
Massachusetts and the Infrared Processing and Analysis
Center/California Institute of Technology, funded by the National
Aeronautics and Space Administration and the National Science
Foundation. This paper also made use of data products from the
DEep Near-Infrared Survey (DENIS) of the Southern Sky. The DENIS
project is supported, in France by the Institut National des
Sciences de l'Univers, the Education Ministry and the Centre
National de la Recherche Scientifique, in Germany by the State of
Baden-W{\rm $\ddot{u}$}rtemberg, in Spain by DGICYT, in Italy by
the Consiglio Nazionale delle Ricerche, in Austria by the Fonds
zur F{\rm $\ddot{o}$}rderung der Wissenschaftlichen Forschung and
the Bundesministerium f{\rm $\ddot{u}$}r Wissenschaft und
Forschung.

We thank the whole ISOGAL team. We also express our gratitude to
Dr. Ian Glass for his careful reading of the manuscript and useful
suggestions. This research is in part supported by the NKBRSF
G19990754 Fund and the NSFC Project 10473003 of China.
\end{acknowledgements}

\clearpage

\begin{footnotesize}
\begin{longtable}[c]{|c|c|c|c|c|c|}
  \caption{Extinctions at 7${\rm \mu}$m from the selected ISOGAL fields }
  \label{tabext7} \\ \hline
    Field Name & Filter & Number & k$_{7}$ & $\sigma_{k_{7}}$ & A$_{7}$ \\ \hline
FA+00580+00076 & 2 & 155 &  0.37 &  0.012  &  0.44  \\
FA+00712+00068 & 2 &  80 &  0.37 &  0.010  &  0.44  \\
FA+00737$-$00070 & 2 & 287 &  0.34 &  0.009  &0.48   \\
\hline
\end{longtable}
\end{footnotesize}

\begin{footnotesize}
\begin{longtable}[c]{|c|c|c|c|c|c|}
  \caption{Extinctions at 15${\rm \mu}$m from the selected ISOGAL fields }
  \label{tabext15} \\ \hline
Field Name & Filter & Number & k$_{15}$ & $\sigma_{k_{15}}$ &
A$_{15}$ \\ \hline
FC$-$01904+00013 &  3  & 58& 0.41 &  0.022 &0.38 \\
FC+00440$-$00009 &  3  & 36& 0.41 &  0.028 &0.38 \\
FC+01514+00075 &  3    & 51& 0.40 &  0.037 &0.40 \\
 \hline
\end{longtable}
\end{footnotesize}


\begin{thebibliography}{}
\bibitem[Belikov et al., 1999]{Belikov99}Belikov, A., Kharchenko, N., Piskunov, A. \&
Schilbach, E. 1999, A\&AS 134, 525
\bibitem[Benjamin et al., 2003]{Benjamin03}
 Benjamin, R., Churchwell, E., Babler, B., et al., 2003, PASP 115, 953
\bibitem[Bertelli et al., 1994]{Bertelli94} Bertelli, G., Bressan, A., Chiosi, C.,
Fagotto, F. \&  Nasi, E. 1994, A\&AS 106, 275
\bibitem[Blommaert et al., 2003]{Blommaert03} Blommaert, J.,
Siebenmorgen, R., Coulais, A., Metcalfe, L., Miville-Deschenes,
M., Okumura, K., Ott, S., Pollock, A., Sauvage, M. \& Starck, J.
2003, ESA SP 1262, Vol.II, p.14
\bibitem[Bronfman et al., 1989]{Bronfman89} Bronfman, L., Alvarez, H., Cohen,
R.S. \& Thaddeus P. 1989, ApJS 71, 481
\bibitem[Cardelli et al., 1989]{Cardelli89} Cardelli, J., Clayton, G. \& Mathis, J.
1989, ApJ 345, 245
\bibitem[Churchwell et al., 2004]{Churchwell04} Churchwell, E., Whitney, B., Babler, B., Indebetouw, R., Meade,
M., Watson, C., Wolff, M., Wolfire, M. \& Bania, T., et al. 2004,
ApJS 154, 322
\bibitem[Combes, 1991]{Combes91} Combes, F. 1991, ARA\&A 29, 195
\bibitem[Cutri et al., 2003]{Cutri03} Cutri, C., Skrutskie, M. \& van
Dyk, S. 2003, available online at
http://www.ipac.caltech.edu/2mass/
\bibitem[Dame et al., 1987]{Dame87} Dame, T., Ungerechts, R., Cohen, R., De Geus, J.,
 et al. 1987, ApJ 322, 706
\bibitem[Draine \& Lee, 1984]{Draine84} Draine, B., \& Lee, H. 1984,
ApJ 285, 89
\bibitem[Draine, 1989]{Draine89} Draine, B. 1989, in Infrared
Spectroscopy in Astronomy, ed. B.H. Kaldeich (Paris, ESA), 93
\bibitem[Drain, 2003]{Draine03} Draine, B. 2003, ARA\&A 41, 241
\bibitem[Epchtein et al., 1994]{Epchtein94} Epchtein, N., de Batz,
B., Copet, E., et al. 1994, Ap\&SS 217, 3
\bibitem[Epchtein et al., 1997]{Epchtein97} Epchtein, N., De Batz, B., Capoani, L., et al. 1997, Messenger 87, 27
\bibitem[Epchtein et al., 1999]{Epchtein99} Epchtein, N., Deul, E., Derriere, S.,
Borsenberger, J., Egret, D., Simon, G., Alard, C., Bal\^{a}zs, L.,
de Batz, B., Cioni, M., et al. 1999, A\&A 349, 236
\bibitem[Felli et al., 2002]{Felli02} Felli, M., Testi, L., Schuller, F., \& Omont, A.
2002, A\&A, 392, 971
\bibitem[Glass, 1999]{Glass99a} Glass, I. 1999, Handbook of Infrared Astronomy, Cambridge, Univ. Press
\bibitem[Glass et al., 1999]{Glass99} Glass, I., Ganesh, S., Alard, C., Blommaert, J., Gilmore, G.,
Lloyd Evans, T., Omont, A., Schultheis, M., Simon, G. 1999, MNRAS
308, 127
\bibitem[Hennebelle et al., 2001]{Hennebelle01} Hennebelle, P.,
P\'{e}rault, M., Teyssier, D., Ganesh, S. 2001, AA, 365, 598
\bibitem[Indebetouw et al., 2005]{Indebetouw04}Indebetouw, R., Mathis, J. S., Babler, B. L., et al., 2005, ApJ 619, 931
\bibitem[Jiang et al., 2003]{Jiang03} Jiang, B.W., Omont, A.,
Ganesh, S., Simon, G., Schuller, F. 2003, AA, 400, 903
\bibitem[Landini et al., 1984]{Landini84} Landini, M., Natta, A.,
Oliva, E., Salinari, P., Moorwood, A. 1984, A\&A 134, 384
\bibitem[Lutz et al., 1996]{Lutz96} Lutz, D., Feuchtgruber, H.,
Genzel, R., Kunze, D. et al. 1996, AA, 315, L269
\bibitem[Lutz, 1999]{Lutz99} Lutz, D. 1999, ESA SP-427: The Universe as Seen By
ISO, 623
\bibitem[Mathis, 1990]{Mathis90} Mathis, J. 1990, ARA\&A 28, 37
\bibitem[Neckel \& Klare, 1980]{Neckel80} Neckel, Th., Klare, G. 1980, A\&AS 42, 251
\bibitem[Omont et al., 1999]{Omont99} Omont, A., Ganesh, S., Alard, C., et al. 1999,
 A\&A 348, 755
\bibitem[Omont et al., 2003]{Omont03} Omont, A., Gilmore, G.,
Alard, C., et al. 2003, AA, 403, 975
\bibitem[Ott et al., 1997]{Ott97} Ott, S., Abergel, A., Altieri, B.,
et al. 1997, Design and Implementation of CIA, the ISOCAM
Interactive Analysis System, in ASP Conf. Ser. 125, ed. Hunt G. \&
Payne H., 34
\bibitem[P\'{e}rault et al., 1996]{Perault96} P\'{e}rault, M., Omont, A., Simon, G., et al. 1996, A\&A 315,
L165
\bibitem[Price et al., 2001]{Price01} Price, S., Egan, M., Carey, S., Mizuno, D.,
 Kuchar, T. 2001, AJ 121, 2819
\bibitem[Rieke \& Lebofsky, 1985]{Rieke85} Rieke, G., \& Lebofsky,
M. 1985, ApJ, 288, 618
\bibitem[Roche \& Aitken, 1985]{Roche85} Roche, P., \& Aitken, D.
1985, MNRAS, 215, 425
\bibitem[Rosenthal et al., 2000]{Rosenthal00} Rosenthal, D.,
Bertoldi, F., Drapatz, S. 2000 A\&A 356, 705
\bibitem[Schuller et al., 2003]{Schuller03} Schuller, F., Ganesh,
S., Messineo, M., et al. 2003, AA, 403, 955
\bibitem[Skrutskie et al., 1997]{Skrutskie97} Skrutskie, M.,
Schneider, S., Stiening, R., et al. 1997, in The Impact of Large
Scale Near-IR Sky Surveys, eds. F.Garzon, N.Epchtein,
A.Omont,B.Burton, \& P.Persi (Netherlands:Kluwer), 187
\bibitem[van der Hucht et al., 1996]{Hucht96} van der Hucht, K., Morris, P., Williams,
P., et al. 1996, AA, 315, L193
\bibitem[van der Hulst, 1946]{Hulst46} van der Hulst, H.C. 1946,
Rech. Astron. Obs. Utrecht, 11, 1
\bibitem[van Loon et al., 2003]{vanLoon03} van Loon, J., Gilmore,
G., Omont, A., Blommaert, J., Glass, I., Messineo, M., Schuller,
F., Schultheis, M., Yamamura, I., Zhao, H.S. 2003, MNRAS, 338, 857
\bibitem[Wainscoat et al., 1992]{Wainscoat92} Wainscoat, R., Cohen, M., Volk, K.,
Walker, H., Schwartz, D.,1992, ApJSS 83, 146
\bibitem[Weingartner \& Draine, 2001]{Weingartner01} Weingartner,
J. \& Draine B.T. 2001, AA, 548, 296
\bibitem[Whittet, 1988]{Whittet88} Whittet, D., 1988, in Dust in
the Universe, eds. Bailey, M. \& Williams, D., p.25

\end{thebibliography}
\end{document}